\documentclass[twocolumn]{autart}    

\usepackage{graphicx}          
\usepackage{graphics}
\usepackage{epsfig}

\usepackage{amssymb}

\usepackage{enumerate}
\usepackage{color}
\usepackage{cases}
\newtheorem{assumption}{Assumption}
\newtheorem{lemma}{Lemma}

\newtheorem{theorem}{Theorem}
\newtheorem{remark}{Remark}

\newtheorem{corollary}{Corollary}         
\begin{document}

\begin{frontmatter}
\title{Distributed Average Tracking for {Multiple Signals Generated
by Linear Dynamical Systems}: An Edge-based Framework} 
\thanks[footnotemark]{Corresponding author }
\author[Nwpu] {Yu Zhao \thanksref{footnotemark}}\ead{yuzhao5977@gmail.com},
\author[Nwpu] {Yongfang Liu},
\author[pku] {Zhongkui Li},            
\author[pku] {Zhisheng Duan}  
\address[Nwpu]{Department of Traffic and Control Engineering, School of Automation, Northwestern Polytechnical University, Xi'an Shaanxi, 710129, China}
\address[pku] {State Key Laboratory for Turbulence and Complex Systems, Department of Mechanics and Engineering Science, College of Engineering, Peking University, Beijing 100871, China}  



\begin{abstract}                          
This paper studies the distributed average tracking problem for multiple time-varying signals {generated
by} linear dynamics, whose reference inputs are nonzero and not available to any agent in the network. In {the} edge-based framework, a pair of continuous algorithms with, respectively, static and adaptive coupling strengths are designed. Based on the boundary layer concept, the proposed continuous algorithm with static coupling strengths can asymptotically track the average of multiple reference signals without the chattering phenomenon. Furthermore, for the case of algorithms with adaptive coupling strengths, average tracking errors are uniformly ultimately bounded and exponentially converge to a small adjustable bounded set. Finally, a simulation example is presented to show the validity of theoretical results.
\end{abstract}

\end{frontmatter}

\section{Introduction}
{In the past two decades, there {have} been lots of interests in the distributed cooperative control \cite{OlfatiSaber1}, \cite{Ren:07}, \cite{Hong:08}, \cite{Cao:12}, \cite{Litac}, \cite{Tuna:2008}, \cite{Zhang:11}, \cite{Liu:15}, \cite{Liu:16}, \cite{Zhao:16}, \cite{Zhao:15}, \cite{Zhao:15scl}, and \cite{Ji}, for multi-agent systems due to its {potential} applications in formation flying, path planning and so forth. {Besides, the clock synchronization problems were also discussed in \cite{Mills91,Sundararaman05,Mauro12,Carli14,Bolognani16}, which are very important to design distributed algorithms.}
Distributed average tracking, as a generalization of consensus and cooperative tracking {problems}, has received increasing attentions and been applied in many different perspectives, such as distributed sensor networks \cite{Spanos2}, \cite{Bai2011} and distributed coordination \cite{Yang}, \cite{Sun}. For practical applications, distributed average tracking should be investigated for signals modeled by more and more complex dynamical systems.}

The objective of distributed average tracking problems is to design a distributed algorithm for multi-agent systems to track the average of multiple reference signals.
The motivation of this problem comes from the coordinated tracking for multiple camera systems. Spurred by the pioneering works in \cite{Spanos1}, and \cite{Freeman} on the distributed average tracking via linear algorithms, { real applications of related results can be found} in distributed sensor fusion \cite{Spanos2}, \cite{Bai2011}, and formation control \cite{Yang}. In \cite{Bai}, distributed average tracking problems were investigated by considering the robustness to initial errors in algorithms. The above-mentioned results are important for scientific researchers to build up a general framework to investigate this topic. {However, a common assumption in the above works is that the multiple reference signals are constants \cite{Freeman} or achieving to values  \cite{Spanos1}.} In practical applications, reference signals may be produced by more general dynamics. {For this reason,} a class of nonlinear algorithms were designed in \cite{Nosrati:12} to track multiple reference signals with bounded deviations. Then, based on non-smooth control {approaches}, a couple of distributed algorithms were proposed in \cite{Chengfei:12} and \cite{Chengfei:13} for agents to track arbitrary time-varying reference signals with bounded {deviations} and {bounded second deviations}, respectively. Using discontinuous algorithms, further, \cite{Zhaoyuicca} studied the distributed average tracking problems for multiple signals generated by linear dynamics.

Motivated by the above mentioned observations, this paper is devoted to solving the distributed average tracking problem with continuous algorithms, for multiple time-varying signals generated by general linear dynamical systems, whose reference inputs are assumed to be nonzero and not available to any agent {in networks}. First of all, based on relative states of neighboring agents, a class of distributed continuous control algorithms are proposed and analyzed.  {Then, a novel class of distributed algorithms with adaptive coupling strengths are designed by utilizing an adaptive control technique.}
{Different from \cite{Cao:12} and \cite{Litac}, where the nonlinear signum function was applied to the whole neighborhood (node-based algorithm), the proposed algorithms in this paper are designed along the edge-based framework as in \cite{Chengfei:12}, \cite{Chengfei:13} and \cite{Zhaoyuicca}}.
{Compared with the above existing results, the contributions of this paper are three-fold. First, main results of this paper extend the reference signals which were generated by first and second-order integrators in \cite{Chengfei:12} and \cite{Chengfei:13}, respectively, to signals generated by linear dynamical systems, which can describe more complex signals.   An advantage of edge-based algorithms designed here is that  {they have} a certain symmetry, which is very important to get the average value of multiple signals under an undirected topology.
By utilizing this property, the edge-based algorithms obtained in this
paper successfully solve distributed average tracking problems for
multiple signals generated by general linear systems with bounded
inputs. Second, by using adaptive control approaches, the requirements of all  global information  {are} removed, which greatly reduce the computational complexity for large-scale networks.
Third, compared with existing results in \cite{Zhaoyuicca}, new continuous algorithms are redesigned via the boundary layer concept with clock synchronization devices. Since there exist differences between the local times of the agents, which may effect the distributed average tracking result, the clock synchronization is introduced in this paper.
The clock synchronization problem has been solved in many existing papers such as \cite{Mills91,Sundararaman05,Mauro12,Carli14,Bolognani16}.   With the help of the existing results on clock synchronization in \cite{Mills91,Sundararaman05,Mauro12,Carli14,Bolognani16}, the first step before beginning computation
is to set the local clock to synchronize the local times. Thus, the boundary layer concept with clock synchronization devices plays a vital role to reduce { the } chattering phenomenon. Continuous algorithms in this paper is more appropriate for real engineering applications.}

\emph{Notations}: Let $R^n$ and $R^{n\times n}$ be sets of real numbers and real matrices, respectively. $I_n$ represents the identity matrix of dimension $n$. Denote by $\mathbf{1}$ a column vector with
all entries equal to one. The matrix inequality $A> (\geq) B$ means that
$A-B$ is positive (semi-) definite. Denote by $A\otimes B$ the Kronecker product
of matrices $A$ and $B$. For a vector $x=(x_1,x_2,\cdots,x_n)^T\in R^n$, let $\|x\|$
denote the 2-norm of $x$, $\mathrm{sig}^{\frac{1}{2}}(x)=(\mathrm{sig}^{\frac{1}{2}}(x_1),\mathrm{sig}^{\frac{1}{2}}(x_2),
\cdots,\mathrm{sig}^{\frac{1}{2}}(x_n))^T$. For a set $V$, $|V|$ represents the number of elements in $V$.

\section{Preliminaries}

\subsection{Graph Theory}
An undirected (simple) graph $\mathcal{G}$ is specified by a vertex set $\mathcal{V}$ and an edge set $\mathcal{E}$ whose elements characterize the incidence relations between distinct pairs of $\mathcal{V}$. The notation $i\sim j$ is used to denote that node $i$ is connected to node $j$, or equivalently, $(i, j)\in \mathcal{E}$. We make use of the $|\mathcal{V}|\times|\mathcal{E}|$ incidence matrix, $D(\mathcal{G})$, for a graph with an arbitrary orientation, i.e., a graph whose edges have a head (a terminal node) and a tail (an initial node). The columns of $D(\mathcal{G})$ are then indexed by the edge set, and the $i$th row entry takes the value $1$ if it is the initial node of the corresponding edge, $-1$ if it is the terminal node, and zero otherwise. The diagonal matrix $\Delta(\mathcal{G})$ of the graph contains the degree of each vertex on its diagonal. The adjacency matrix, $A(\mathcal{G})$, is the $|\mathcal{V}|\times|\mathcal{V}|$ symmetric matrix with zero in the diagonal and one in the $(i,j)$th position if node $i$ is adjacent to node $j$. The graph Laplacian  \cite{GraphTheory} of $\mathcal{G}$, $L:= {\frac{1}{2}}D(\mathcal{G})D(\mathcal{G})^T=\Delta(\mathcal{G})-A(\mathcal{G})$,
is a rank deficient positive semi-definite matrix.

An undirected path between node $i_1$ and node $i_s$ on undirected graph means a sequence of ordered undirected edges with the form $(i_k; i_{k+1}), k = 1, \cdots, s-1$.
A graph $\mathcal{G}$ is said to be connected if there exists a path between each pair of distinct nodes.

\begin{assumption}\label{ass}
Graph $\mathcal{G}$ is undirected and connected.
\end{assumption}

\begin{lemma} \label{lemma1}\cite{GraphTheory}
Under Assumption \ref{ass}, zero is a simple eigenvalue of $L$ with
$\mathbf{1}$ as an eigenvector and all the other eigenvalues are positive. Moreover, the
smallest nonzero eigenvalue $\lambda_2$ of L satisfies $\lambda_2=\min\limits_{ x\neq 0, \mathbf{1}^Tx=0 } \frac{x^TLx}{x^Tx}$.
\end{lemma}

{Define $M=I_N-\frac{1}{N}\mathbf{1}\mathbf{1}^T$. Then $M$ satisfies following properties: Firstly, it is easy to see that $0$ is a simple eigenvalue
of $M$ with $\mathbf{1}$ as {the} corresponding right eigenvector and $1$ is the other eigenvalue with multiplicity $N-1$, i.e., $M\mathbf{1}=\mathbf{1}^TM=0$. Secondly, since $L^T=L$, one has $LM=L(I_N-\frac{1}{N}\mathbf{1}\mathbf{1}^T)=L-\frac{1}{N}L\mathbf{1}\mathbf{1}^T=L=L-\frac{1}{N}\mathbf{1}\mathbf{1}^TL=(I_N-\frac{1}{N}\mathbf{1}\mathbf{1}^T)L=
ML$. Finally, $M^2=M(I_N-\frac{1}{N}\mathbf{1}\mathbf{1}^T)=M-\frac{1}{N}M\mathbf{1}\mathbf{1}^T=M$.
}

\section{Distributed average tracking for multiple reference signals with general linear dynamics}

Suppose that there are $N$ time-varying reference signals, $r_i(t)\in R^n, i=1,2,\cdots, N$, which generated by  the following linear dynamical systems:
\begin{eqnarray}\label{L referencesignals}
\dot{r}_i(t)=Ar_i(t)+Bf_i(t),
\end{eqnarray}
{where $A\in R^{n\times n}$ and $B\in R^{n\times p}$ both are constant matrices with compatible dimensions,  {$r_i(t)\in R^{n}$} is the state of the $i$th signal, and $f_i(t)\in R^{p} $} represents the reference input of the $i$th signal.  Here, we assume that $f_i(t)$ is continuous and bounded,  i.e., $\|f_i(t)\|\leq f_0$, for $i=1,2,\cdots, N$, where $f_0$ is a positive constant. Suppose that there are $N$ agents with $x_i\in R^n$ being the state of the $i$th agent in distributed algorithms. It is assumed that agent $i$ has access to $r_i(t)$, and agent $i$ can obtain the relative information from its neighbors denoted by $\mathcal{N}_i$.  {Besides, let $|\mathcal{N}_i|$ represent the number of  elements in the set $\mathcal{N}_i$, $i=1,2,\cdots, N$.}
 {\begin{assumption}\label{ass2}
$(A,B)$ is stabilizable.
\end{assumption}}
The main objective of this paper is to design a class of distributed algorithms for agents to track the average of multiple signals $r_i(t)$ generated by the general linear dynamics  (\ref{L referencesignals}) with bounded reference inputs $f_i(t),\;i=1,2,\cdots,N$.

Therefore, a distributed algorithm is proposed as follows:
\begin{eqnarray}\label{L distributed control algorithm}
\dot{s}_i(t) &=&As_i(t)+Bu_i(t),\nonumber\\
u_i(t)&=&c_1 \sum_{j\in \mathcal{N}_i}[K(x_i(t)-x_j(t))]\nonumber\\
&&+c_2 \sum_{j\in \mathcal{N}_i} h_i [K(x_i(t)-x_j(t)), t_i],\nonumber \\
{x}_i(t) &=& s_i(t)+r_i(t),
\end{eqnarray}
where $s_i(t), \;i=1,2,\cdots,N$, are internal states of the distributed filter (\ref{L distributed control algorithm}), $c_1$, $c_2$ and $K$ are coupling strengths and feedback gain matrix, respectively, to be determined. Besides, the nonlinear function $h_i(\cdot)$ is defined as follows: for $\omega\in R^n$,
\begin{eqnarray}\label{hi}
h_i(\omega,  {t_i})=\frac{\omega}{\|\omega\|+\varepsilon e^{-\varphi {t_i}} },
\end{eqnarray}
with a finite-time clock synchronization device:
\begin{eqnarray}\label{time}
  \frac{dt_i}{dt} &=& 1+\sum_{j\in \mathcal{N}_i}\mathrm{sig}^{\frac{1}{2}} (t_i-t_j), \;\;\;\;\; i=1,2,\cdots,N,
\end{eqnarray}
where $\varepsilon$ and $\varphi$ are positive constants, $t_i$ is a local time of the local clock in agent $i$.

 {{\remark{
Besides (\ref{time}), there exist many algorithms on designing clock synchronization device in  \cite{Mills91,Sundararaman05,Mauro12,Carli14,Bolognani16}, where the robust finite-time clock synchronization device is considered in \cite{Mauro12}.
Thus, by using
the nonlinear function as given in (\ref{hi}) with (\ref{time}),
the first step before beginning computation is to set the local clock by using the  clock synchronization device (\ref{time}) such that all local times $t_i$ to be identical in a finite settling time $t_0$.
Then, the Algorithm (\ref{L distributed control algorithm}) can
solve the distributed average tracking problem without errors.}}
{\remark{In practice,  there exists external disturbance, which may result the failure of the clock
synchronization device (\ref{time}). For this case, we can use the nonlinear function (\ref{hi}) with $\varphi=0$.
Then, the Algorithm (\ref{L distributed control algorithm}) can solve the distributed average tracking problems with a uniformly ultimately bounded error. As well known, the bounded result is significant in real application.}
}
}

Note that the nonlinear function  $h_i(\omega, t_i)$ in (\ref{hi}) is continuous,
which is actually continuous approximations, via the boundary layer concept \cite{28}, of the discontinuous function
\begin{eqnarray*}
\widehat{h}_i(\omega)= \Bigg\{\begin{array}{cc}
                                                             \frac{\omega}{\|\omega\|} & \;\;\mathrm{if} \;\; \omega\neq 0, \\
                                                             0 & \;\;\mathrm{if} \;\;\omega= 0.\\
                                                           \end{array}
\end{eqnarray*}
The item $\varepsilon e^{-\varphi t_i}$ in (\ref{hi}) defines the size of the boundary layer. As $t_i\rightarrow \infty$, the continuous function $h_i(\omega, t_i)$ approaches the discontinuous function $\widehat{h}_i(\omega)$.

It follows from (\ref{L referencesignals}) and (\ref{L distributed control algorithm}) that the closed-loop system is described by
\begin{eqnarray}\label{L closedloop}
\dot{x}_i(t)&=&Ax_i(t)+c_1B\sum_{j\in \mathcal{N}_i}[K(x_i(t)-x_j(t))]\nonumber\\
&&+c_2B\sum_{j\in \mathcal{N}_i} h_i [K(x_i(t)-x_j(t)), t_i]+Bf_i(t).
\end{eqnarray}

Before moving on, an important lemma is proposed.
\begin{lemma}\label{L lemma3}
Under Assumption \ref{ass}, states $x_i(t)$ in (\ref{L distributed control algorithm}) with  $s_i(t_0)=0$ will track the average of multiple signals, i.e., $\|x_i(t)-\frac{1}{N}\sum_{k=1}^Nr_k(t)\|=0$, if the closed-loop system (\ref{L closedloop}) achieves consensus, i.e., $\lim_{t\rightarrow \infty}\|x_i-\frac{1}{N}\sum_{k=1}^N x_k\|=0$  for $i=1,2,\cdots, N$.
\end{lemma}
\textbf{Proof}: It follows from Assumption \ref{ass} and Remark 1 that
\begin{eqnarray}\label{L1}
&& \sum_{i=1}^N\sum_{j\in \mathcal{N}_i}[K(x_i(t)-x_j(t))]=0,\nonumber\\
&& \sum_{i=1}^N\sum_{j\in \mathcal{N}_i} h_i [K(x_i(t)-x_j(t)), t_i]=0,\;\; t\geq t_0.
\end{eqnarray}
 {Let $s(t)=\sum_{i=1}^Nx_i(t)-\sum_{i=1}^Nr_i(t)$.  From (\ref{L referencesignals}), (\ref{L closedloop}) and (\ref{L1}), we have
\begin{eqnarray}\label{L2}
\dot{s}(t)=As(t),\;\; t\geq t_0,
\end{eqnarray}
with $s(t_0)=0$.
By solving the differential equation (\ref{L2}) with initial condition above, we always have $\lim_{t\to \infty}s(t)=\lim_{t\to \infty}e^{A(t-t_0)}s(t_0)=0$.
Thus, we obtain
\begin{eqnarray}\label{L3}
\lim_{t\to \infty}\sum_{i=1}^Nx_i(t)=\lim_{t\to \infty}\sum_{i=1}^Nr_i(t).
\end{eqnarray}}
According to  Assumption \ref{ass}, if $x_i(t)$ in (\ref{L closedloop}) achieves consensus, i.e., $\lim_{t\rightarrow \infty}\|x_i(t)-\frac{1}{N}\sum_{k=1}^Nx_k(t)\|=0$ for $i=1,2,\cdots, N$, then, it follows from (\ref{L3}) that $\lim_{t\rightarrow \infty}\|x_i-\frac{1}{N}\sum_{k=1}^Nr_k(t)\|=0$, for $i=1,2,\cdots, N$. This completes the proof.
\begin{remark}
In the proof of Lemma \ref{L lemma3}, it requires that $ s_i(t_0)= 0$, which is a necessary condition to draw conclusions, when $A$ is not asymptotically stable. In the case that $A$ is asymptotically stable, without requiring the initial condition  $ s_i(t_0)= 0$, we can still reach the same conclusions as shown in Lemma \ref{L lemma3}, since the solution of (\ref{L2}) will converge to the origin for any initial condition.
\end{remark}

Let $x(t)=(x_1^T(t),x_2^T(t),\cdots,x_N^T(t))^T$, and $F(t)=(f_1^T(t),f_2^T(t),\cdots,f_N^T(t))^T$.
Define $\xi(t)=(M\otimes I)x(t)$, where $\xi(t)=(\xi_1^T(t),\xi_2^T(t),\cdots,\xi_N^T(t))^T$.
Then, it follows that $\xi(t)= 0$ if and only if $x_1(t)=x_2(t)=\cdots=x_N(t)$. Therefore, the consensus problem of (\ref{L closedloop}) is solved if and only if $\xi(t)$ asymptotically converges to zero. Hereafter, we refer to $\xi(t)$ as the consensus error. By
noting that $LM = L$ and $MD(\mathcal{G})=D(\mathcal{G})$, it is not difficult to obtain from (\ref{L closedloop}) that the consensus error $\xi(t)$ satisfies
{\begin{eqnarray}\label{LME closedloop}
\dot{\xi}(t)&=&(I\otimes A)\xi(t)+c_1(L\otimes BK)\xi(t)\nonumber\\
&&+c_2\left[
                                              \begin{array}{c}
                                                B\sum\limits_{j\in \mathcal{N}_1}h_1 [K(\xi_1(t)-\xi_j(t)), t_1] \\
                                                \vdots \\
                                                B\sum\limits_{j\in \mathcal{N}_N}h_N [K(\xi_N(t)-\xi_j(t)), t_N] \\
                                              \end{array}
                                      \right]\nonumber\\
&&+(M\otimes B)F(t).
\end{eqnarray}}

\textbf{Algorithm 1:} {Under Assumptions 1 and 2,} for multiple reference signals in (\ref{L referencesignals}), the distributed average tracking algorithm (\ref{L distributed control algorithm}) can be constructed as follows
\begin{enumerate}
{\item Set the local clock  such that the synchronization of the local time  $t_i$ in finite time by using the clock synchronization device (\ref{time}).}
  \item Solve the algebraic Ricatti equation (ARE):
\begin{eqnarray}\label{LMI}
PA+A^TP-PBB^TP+Q=0,
\end{eqnarray}
with $Q>0$ to obtain a matrix $P>0$. Then, choose $K=-B^TP$.
  \item Select the first coupling strength $c_1\geq \frac{1}{2\lambda_{2}}$, where $\lambda_{2}$ is the smallest nonzero eigenvalue of the Laplacian $L$ of $\mathcal{G}$.
  \item Choose the second coupling strength $c_2\geq f_0(N-1)$, where $f_0$ is defined as in (\ref{L referencesignals}).
\end{enumerate}

 {It is worthwhile to mention that the
originality of the Riccati based approach in step (2) in Algorithm 1 for the design of  matrix $K$ can be found  in \cite{Tuna:2008} and \cite{Zhang:11}.}

\begin{theorem}\label{L theorem1f}
Under Assumptions 1 and 2, the state $x_i(t)$ in (\ref{L distributed control algorithm}) will track the average of multiple reference signals $r_i(t),\;i=1,2,\cdots, N$, generated by the general linear dynamics (\ref{L referencesignals}) with bounded reference inputs if coupling strengths $c_1$, $c_2$ and the feedback gain $K$  are designed by Algorithm 1.
\end{theorem}
\textbf{Proof}:
Consider the Lyapunov function candidate
\begin{eqnarray}\label{LV 1}
V_1(t)= \xi^T(M\otimes P)\xi.
\end{eqnarray}
By the definition of $\xi(t)$, it is easy to see that $(\mathbf{1}^T\otimes I)\xi = 0$.
For the connected graph $\mathcal{G}$, it then follows from Lemma \ref{lemma1} that
\begin{eqnarray}\label{gLV 1}
V_1(t)\geq \lambda_{\min}(P)\|\xi\|^2,
\end{eqnarray}
where $\lambda_{\min}(P)$ is the smallest eigenvalue of the positive matrix $P$.
The time derivative of $V_1$ along (\ref{LME closedloop}) can be obtained as follows
\begin{eqnarray}\label{dLV1}
\dot{ {V}}_1&=& \dot{\xi}^T(M\otimes P)\xi+ \xi^T(M\otimes P)\dot{\xi} \nonumber\\
&=& \xi^T(I\otimes A^T+c_1L\otimes K^TB^T)(M\otimes P)\xi\nonumber\\
&&+\xi^T(M\otimes P)(I\otimes A+c_1L\otimes BK)\xi\nonumber\\
&&+2 c_2\xi^T \left[
                                              \begin{array}{c}
                                                PB\sum\limits_{j\in \mathcal{N}_1}h_1 [K(\xi_1(t)-\xi_j(t)), t_1] \\
                                                \vdots \\
                                                PB\sum\limits_{j\in \mathcal{N}_N}h_N [K(\xi_N(t)-\xi_j(t)), t_N] \\
                                              \end{array}
                                      \right]\nonumber\\
&&+2\xi^T(M\otimes PB)F(t).
\end{eqnarray}
Substituting $K=-B^TP$ into (\ref{dLV1}), it follows from the fact $LM=ML=L$ that
\begin{eqnarray}\label{ddLV1}
\dot{ {V}}_1
&=& \xi^T(M\otimes (PA+A^TP)-2c_1L\otimes PBB^TP)\xi\nonumber\\
&&-2 c_2\xi^T \left[
                                              \begin{array}{c}
                                                PB\sum\limits_{j\in \mathcal{N}_1}h_1 [B^TP(\xi_1(t)-\xi_j(t)), t_1] \\
                                                \vdots \\
                                                PB\sum\limits_{j\in \mathcal{N}_N}h_N [B^TP(\xi_N(t)-\xi_j(t)), t_N] \\
                                              \end{array}
                                      \right]\nonumber\\
&&+2\xi^T(M\otimes PB)F(t).
\end{eqnarray}
Since $\|F\|  \leq \sqrt{N}f_0$, we have
\begin{eqnarray}\label{lf5}
&&\xi^T(M\otimes PB)F(t)\nonumber\\
&\leq& \|(M\otimes B^TP)\xi\| \|F(t)\| \nonumber\\
&\leq&
\frac{f_0}{\sqrt{N}}\sum_{i=1}^N\sum_{j=1,j\neq i}^N\|B^TP(\xi_i-\xi_j)\|\nonumber\\
&\leq& \frac{f_0}{\sqrt{N}}\sum_{i=1}^N \max_i\bigg\{\sum_{j=1,j\neq i}^N\|B^TP(\xi_i-\xi_j)\|\bigg\}\nonumber\\
&=& \sqrt{N}f_0 \max_i\bigg\{\sum_{j=1,j\neq i}^N\|B^TP(\xi_i-\xi_j)\|\bigg\}\nonumber\\
&\leq& \frac{f_0}{2}(N-1)\sqrt{N} \sum_{i=1}^N\sum_{j\in \mathcal{N}_i}
\|B^TP(\xi_i-\xi_j)\|.
\end{eqnarray}
Then, because of the fact that $\omega^Th_i(\omega, t_i)= \frac{\|\omega\|^2}{\|\omega\|+\varepsilon e^{-\varphi {t_i}}}$, we get
\begin{eqnarray}\label{lf6}
&&-2c_2\xi^T \left[
                                              \begin{array}{c}
                                                PB\sum\limits_{j\in \mathcal{N}_1}h_1 [B^TP(\xi_1(t)-\xi_j(t)), t_1] \\
                                                \vdots \\
                                                PB\sum\limits_{j\in \mathcal{N}_N}h_N [B^TP(\xi_N(t)-\xi_j(t)), t_N] \\
                                              \end{array}
                                      \right]\nonumber\\
&=&-c_2 \sum_{i=1}^N\sum_{j\in \mathcal{N}_i} \frac{\|B^TP(\xi_i-\xi_j)\|^2}{\|B^TP(\xi_i-\xi_j)\|+\varepsilon e^{-\varphi {t_i}}}.
\end{eqnarray}
By combining with (\ref{lf5}) and (\ref{lf6}), it follows from (\ref{ddLV1}) that
\begin{eqnarray}\label{dddLV1}
\dot{ {V}}_1
&\leq&\xi^T(M{\otimes} (PA{+}A^TP){-}2c_1L{\otimes} PBB^TP)\xi\nonumber\\
&&+f_0(N-1) {\sqrt{N}} \sum_{i=1}^N\sum_{j\in \mathcal{N}_i} \|B^TP(\xi_i-\xi_j)\|\nonumber\\
&&-c_2 \sum_{i=1}^N\sum_{j\in \mathcal{N}_i}
\frac{\|B^TP(\xi_i-\xi_j)\|^2}{\|B^TP(\xi_i-\xi_j)\|+\varepsilon e^{-\varphi {t_i}}}.
\end{eqnarray}
Choose $c_2\geq f_0 (N-1)\sqrt{N}$. We have
\begin{eqnarray}\label{ddddLV1}
\dot{ {V}}_1
&\leq&\xi^T(M {\otimes} (PA{+}A^TP){-}2c_1L{\otimes} PBB^TP)\xi\nonumber\\
&&+c_2 \sum_{i=1}^N\sum_{j\in \mathcal{N}_i}
\Big(\|B^TP(\xi_i-\xi_j)\|\nonumber\\
&&-\frac{\|B^TP(\xi_i-\xi_j)\|^2}{\|B^TP(\xi_i-\xi_j)\|+\varepsilon e^{-\varphi {t_i}}}\Big)\nonumber\\
&\leq&\xi^T(M {\otimes} (PA{+}A^TP){-}2c_1L{\otimes} PBB^TP)\xi\nonumber\\
&&+c_2 \sum_{i=1}^N  {|\mathcal{N}_i|}
\varepsilon e^{-\varphi  {t_i}}.
\end{eqnarray}
 {By Assumption \ref{ass}, there exists an unitary matrix $U$ such that $L=U^T\Lambda U$,} where $\Lambda=\mathrm{diag}(\lambda_1,\lambda_2,\cdots,\lambda_N)$. Without loss of generality, assume that $0=\lambda_1<\lambda_2\leq\cdots\leq\lambda_N$. Thereby, following from the fact that $M^2=M$, we obtain
\begin{eqnarray}\label{dddddLV1}
&&\xi^T(M {\otimes} (PA{+}A^TP){-}2c_1L{\otimes} PBB^TP)\xi\nonumber\\
&=&\xi^T(MU^T\otimes I) [I\otimes (PA+A^TP)\nonumber\\
&&-2c_1\Lambda\otimes PBB^TP](U M\otimes I)\xi\nonumber\\
&\leq&\xi^T(M \otimes (PA+A^TP-2c_1\lambda_2 PBB^TP))\xi.
\end{eqnarray}
Select $c_1\geq \frac{1}{2\lambda_2}$. It follows from (\ref{LMI}) that $PA+A^TP-2c_1\lambda_2 PBB^TP\leq-Q$. Therefore, we have
\begin{eqnarray}\label{ddddddLV1}
\dot{ {V}}_1&<&- \gamma V_1+c_2 \sum_{i=1}^N |\mathcal{N}_i|\varepsilon e^{-\varphi {t_i}},
\end{eqnarray}
where $\gamma=\frac{\lambda_{\min}(Q)}{\lambda_{\max}(P)}$. Thus, we obtain that
\begin{eqnarray*}\label{ddddddoV1}
0{\leq}{ {V}}_1(t) {\leq} e^{-\gamma t}V_1(t_0){+}c_2\sum_{i=1}^N |\mathcal{N}_i|\int_{t_0}^t\varepsilon e^{-\gamma(t-\tau){-}\varphi (\tau+ \eta)}d\tau,\end{eqnarray*} where $\eta=\pi (t_0)$ is a constant.
By noting that
\begin{eqnarray*}\label{dddddddoV1}
&&\int_{t_0}^t\varepsilon e^{-\gamma(t-\tau)-\varphi \tau}d\tau  \\
&=& \Bigg\{\begin{array}{cc}
                                                             \varepsilon (t-t_0)e^{-\gamma t} & \;\;\mathrm{if} \;\; \gamma=\varphi, \\
                                                             \frac{\varepsilon}{\gamma-\varphi}(e^{-(\varphi+\gamma) t}-e^{-\gamma t-\varphi t_0}) & \;\;\mathrm{if} \;\;\gamma \neq \varphi,
                                                           \end{array}
\end{eqnarray*}
we have that $V_1(t)$ will converge to {the} origin as $t\rightarrow \infty$, which means that states of (\ref{L closedloop}) will achieve consensus. Then, according to Lemma \ref{L lemma3}, we have that  tracking errors $\xi_i, i=1,2,\cdots,N$ satisfy
$
\lim_{t\rightarrow \infty}\xi_i(t)=\lim_{t\rightarrow \infty}\bigg(x_i(t)-\frac{1}{N}\sum_{k=1}^{N}x_k(t)\bigg)
=\lim_{t\rightarrow \infty}\bigg(x_i(t)-\frac{1}{N}\sum_{k=1}^{N}r_k(t)\bigg)
=0.
$
Therefore, the distributed average tracking problem is solved. This completes the proof.
 {{\remark{As mentioned in Remark 1, for the case with external disturbance, let $\varphi=0$. Therefore, the nonlinear function (\ref{hi}) is reduced to $h_i(\omega)=\frac{\omega}{\|\omega\|+\varepsilon}$. From the (\ref{ddddddLV1}), one has
$
\dot{ {V}}_1<- \gamma V_1+c_2 \sum_{i=1}^N |\mathcal{N}_i|\varepsilon.
$
Then, the tracking error $\xi$ given in  {(\ref{LME closedloop})} uniformly ultimately
bounded. According to (\ref{gLV 1}), $\xi$  will exponentially converge to the following set
     $
\Omega_0\triangleq \bigg\{\xi: \|\xi\|<\bigg(\frac{c_2 }{\gamma\lambda_{min}(P)}\sum_{i=1}^N |\mathcal{N}_i|\varepsilon\bigg)^\frac{1}{2}\bigg\}.
$
The bounded result is meaningful in real application.
}}
}

 {
In distributed algorithm (\ref{L distributed control algorithm}), it requires the initial state of $s_i(t)$ satisfying $x_i(t_0)=r_i(t_0)$. In order to remove the initial condition $x_i(t_0)=r_i(t_0)$, a modified algorithm is proposed as follows:
\begin{eqnarray}\label{N distributed control algorithm}
\dot{s}_i(t)&=&As_i(t)+Bu_i(t),\nonumber \\
u_i(t)&=&Kx_i(t)+c_2\sum_{j\in \mathcal{N}_i} h_i [K(x_i(t)-x_j(t)), t_i],\nonumber \\
{x}_i(t) &=& s_i(t)+r_i(t).
\end{eqnarray}
\begin{corollary}
By using the modified robust algorithm (\ref{N distributed control algorithm}) with steps (1), (2) and (4) in Algorithm 1, the distributed average tracking can be achieved without requiring {the} initial condition $x_i(t_0)=r_i(t_0)$.
\end{corollary}
\textbf{Proof}: First of all,
let $x_i(t)=s_i(t)+r_i(t)$. From (\ref{N distributed control algorithm}), one has the closed-loop system of (\ref{L referencesignals}) and (\ref{N distributed control algorithm}) is described by
\begin{eqnarray}\label{N closedloop}
\dot{x}_i(t)&=&(A+BK)x_i(t)\nonumber\\
&&+c_2B\sum_{j\in \mathcal{N}_i} h_i [K(x_i(t)-x_j(t)), t_i] +Bf_i(t).
\end{eqnarray}
Then,  in the matrix form, let $\xi=(M\otimes I) x$. The error system is given as follows:
\begin{eqnarray}\label{NLME closedloop}
\dot{\xi}(t)&=&[I\otimes (A+BK)]\xi(t)\nonumber\\
&&-c_2\left[
                                              \begin{array}{c}
                                                B\sum\limits_{j\in \mathcal{N}_1}h_1 [K(\xi_1(t)-\xi_j(t)), t_1] \\
                                                \vdots \\
                                                B\sum\limits_{j\in \mathcal{N}_N}h_N [K(\xi_N(t)-\xi_j(t)), t_N] \\
                                              \end{array}
                                      \right]\nonumber\\
&&+(M\otimes B)F(t).
\end{eqnarray}
Consider the same Lyapunov function candidate in (\ref{LV 1}). One has
\begin{eqnarray}\label{NddOV1}
\dot{ {V}}_1
&=& \xi^T[M \otimes (PA+A^TP-PBB^TP)]\xi\nonumber\\
&&-2 c_2\xi^T \left[
                                              \begin{array}{c}
                                                PB\sum\limits_{j\in \mathcal{N}_1}h_1 [B^TP(\xi_1(t)-\xi_j(t)), t_1] \\
                                                \vdots \\
                                                PB\sum\limits_{j\in \mathcal{N}_N}h_N [B^TP(\xi_N(t)-\xi_j(t)), t_N] \\
                                              \end{array}
                                      \right]\nonumber\\
&&+2\xi^T(M\otimes PB)F(t).
\end{eqnarray}
Similar to the proof of (\ref{lf5})-(\ref{ddddLV1}) in Theorem \ref{L theorem1f}, one has
\begin{eqnarray}\label{NdddOV1}
\dot{ {V}}_1
&=& \xi^T[M \otimes (PA+A^TP-PBB^TP)]\xi \nonumber\\
&&+c_2 \sum_{i=1}^N|\mathcal{N}_i|
\varepsilon e^{-\varphi  {t_i}}.
\end{eqnarray}
From (\ref{LMI}), one obtains the same result in Theorem \ref{L theorem1f}. Thus, $\lim_{t\rightarrow \infty}\|x_i(t)-\frac{1}{N}\sum_{k=1}^Nx_k(t)\|=0$ for $i=1,2,\cdots, N$.}

 {
Second, let $s(t)=\sum_{i=1}^Nx_i(t)-\sum_{i=1}^Nr_i(t)$. We have
\begin{eqnarray}\label{LL2}
\dot{s}(t)=(A+BK)s(t), \;\;t>t_0.
\end{eqnarray}
By solving the differential equation (\ref{LL2}) with $A+BK$
 {being
asymptotically stable,} we always have $\lim_{t\rightarrow \infty}s(t)=0$.
Thus, we obtain $
\lim_{t\rightarrow \infty}\sum_{i=1}^N\\x_i(t)=\lim_{t\rightarrow \infty}\sum_{i=1}^Nr_i(t).
$
It follows that
$\lim_{t\rightarrow \infty}\|x_i(t)-\frac{1}{N}\sum_{k=1}^Nr_k(t)\|
=\lim_{t\rightarrow \infty}\|x_i(t)-\frac{1}{N}\sum_{k=1}^Nx_k(t)\|
=0,
$ for $i=1,2,\cdots, N$. This completes the proof.
\begin{remark}
It is worth mentioning that, different from the consensus problem in existing papers \cite{Hong:08}-\cite{Zhang:11}, where algorithms were designed in node-based viewpoints, one advantage of edge-based algorithms designed here is that they have a certain symmetry in networks, which are very important to get the average value of the multiple signals under an undirected topology. By utilizing the symmetry in the edge-based framework, the algorithm (\ref{L distributed control algorithm}) is designed. Different from node-based algorithms in \cite{Hong:08}-\cite{Zhang:11}, which can not solve average tracking problems, the edge-based algorithm in this paper can ensure the state of each agent to track the average value of multiple signals. Besides, in \cite{Litac}, it studied consensus problems of multiple linear systems with discontinuous algorithms. The discontinuous algorithm can not be realized in practical applications for its large chattering. In order to reduce the chattering effect, by using the boundary layer approximation, continuous algorithms  are proposed  in this paper. Compared with the result in \cite{Litac}, the main contribution of this paper lies to the feasibility of continuous algorithms in  {practical applications}.
\end{remark}
}

\section{Distributed average tracking with distributed adaptive coupling strengths}
Note that in the  {above} section, the first coupling strength $c_1$, designed as $c_1>\frac{1}{2\lambda_2 }$,  {depends} on the communication topology. The second coupling strength $c_2$, designed as $c_2>f_0(N-1)\sqrt{N}$, requires $f_0$ and $N$. Generally, the smallest nonzero eigenvalue $\lambda_2$, the number $N$ of vertex set $\mathcal{V}$ and the {upper} bound $f_0$ of $f_i(t)$ all are global information, which are difficult to be obtained by agents when the scale of the network is very large. Therefore, to overcome these restrictions, a distributed average tracking algorithm with distributed adaptive coupling strengths is proposed as follows:
\begin{eqnarray}\label{A distributed control algorithm}
\dot{s}_i(t) &=&As_i(t)+Bu_i(t),\nonumber\\
u_i(t)&=&\sum_{j\in \mathcal{N}_i}\alpha_{ij}(t)[K(x_i(t)-x_j(t))]\nonumber\\
&&+B\sum_{j\in \mathcal{N}_i} \beta_{ij}(t)h_i [K(x_i(t)-x_j(t)), t_i],\nonumber \\
{x}_i(t) &=& s_i(t)+r_i(t),\;s_i(t_0)=0,
\end{eqnarray}
with distributed adaptive laws
\begin{eqnarray}\label{A distributed adaptive law}
\dot{\alpha}_{ij}(t)&=&\mu[-\vartheta \alpha_{ij}(t)\nonumber\\
&&+(x_i(t)-x_j(t))^T \Gamma (x_i(t)-x_j(t))], \nonumber\\
\dot{\beta}_{ij}(t) &=&
\nu\Bigg[-\chi \beta_{ij}(t)\nonumber\\
&&+\frac{\|K(x_i(t)-x_j(t))\|^2}{\|K(x_i(t)-x_j(t))\|+\varepsilon e^{-\varphi {t_i}}}\Bigg],
\end{eqnarray}
where $\alpha_{ij}(t)$ and $\beta_{ij}(t)$ are two adaptive coupling strengths satisfying {${\alpha}_{ij}(t_0)=0$ and ${\beta}_{ij}(t_0)=0$}, $\Gamma\in R^{n\times n}$ is a constant gain matrix, $\mu$, $\nu$, $\vartheta$ and $\chi$ are positive constants.

It follows from (\ref{L referencesignals}) and (\ref{A distributed control algorithm}) that the closed-loop system is described by
\begin{eqnarray}\label{A closedloop}
\dot{x}_i(t)&=&Ax_i(t)+B\sum_{j\in \mathcal{N}_i}\alpha_{ij}(t)[K(x_i(t)-x_j(t))]\nonumber\\
&&{+}B\sum_{j\in \mathcal{N}_i} \beta_{ij}(t)h_i[K(x_i(t){-}x_j(t)), t_i]{+}Bf_i(t),\nonumber\\
\end{eqnarray}
where $\alpha_{ij}(t)$ and $\beta_{ij}(t)$  are given by (\ref{A distributed adaptive law}).

Similarly as in the above section, the following lemma is firstly given.
\begin{lemma}\label{A lemma3}
Under Assumption \ref{ass}, for the algorithm (\ref{A distributed control algorithm}) with (\ref{A distributed adaptive law}), if $\lim_{t\rightarrow 0}\|x_i-\frac{1}{N}\sum_{k=1}^Nx_k\|=0,\;i=1,2,\cdots, N$, then $\lim_{t\rightarrow \infty}\|x_i-\frac{1}{N}\sum_{k=1}^Nr_k\|=0,\;i=1,2,\cdots, N$.
\end{lemma}
\textbf{Proof}: Since $\alpha_{ij}(t_0)=0$ and $\beta_{ij}(t_0)=0$, it follows from (\ref{A distributed adaptive law}) that $\alpha_{ij}(t)=\alpha_{ji}(t)$ and $\beta_{ij}(t)=\beta_{ji}(t)$. From Assumption \ref{ass}, we have
\begin{eqnarray}\label{A1}
&&\sum_{i=1}^N\sum_{j\in \mathcal{N}_i}\alpha_{ij}(t)[K(x_i(t)-x_j(t))]=0,\nonumber\\
&&\sum_{i=1}^N\sum_{j\in \mathcal{N}_i}\beta_{ij}(t) h_i
[K(x_i(t)-x_j(t)), t_i]=0, \;\;t>t_0.\nonumber
\end{eqnarray}
Similar to the proof of Lemma \ref{L lemma3}, we can draw the conclusion in (\ref{L3}).
This completes the proof.

\textbf{Algorithm 2:}
For multiple reference signals in (\ref{L referencesignals}), the distributed average tracking algorithm (\ref{A distributed control algorithm}) with adaptive laws (\ref{A distributed adaptive law}) can be constructed as follows
\begin{enumerate}
\item  {Set the local clock such that the synchronization of the local time $t_i$ in finite time by using the clock synchronization device (\ref{time})}.
  \item Solve the ARE (\ref{LMI}) with $Q>0$
to obtain a matrix $P>0$. Then, choose $\Gamma=PBB^TP$ and $K=-B^TP$, respectively.
  \item Select $\mu$ and $\nu$ small enough, respectively, such that $\varrho \triangleq\max\{\mu\vartheta, \nu\chi\}<\gamma$, where $\gamma=\frac{\lambda_{\min}(Q)}{\lambda_{\max}(P)}$.
\end{enumerate}

The following theorem shows the ultimate boundedness of tracking errors and adaptive coupling strengths.
\begin{theorem}\label{A theorem1f}
Under the Assumption \ref{ass}, the fully distributed average tracking problem is solved by (\ref{A distributed control algorithm}) with (\ref{A distributed adaptive law}) if feedback gains $\Gamma$ and $K$ are designed as given in Algorithm 2. The tracking error $\xi$ defined in  {(\ref{LME closedloop})} and adaptive gains $\alpha_{ij}(t)$ and $\beta_{ij}(t)$ are uniformly ultimately
bounded and  following statements are hold:
\begin{enumerate}
  \item For any $\vartheta$ and $\chi$, $\xi$, $\widetilde{\alpha}_{ij}$ and $\widetilde{\beta}_{ij}$ exponentially converge to the following bounded set
      \begin{eqnarray}\label{omiga1}
\Omega_1&\triangleq &\Bigg\{\xi, \widetilde{\alpha}_{ij}(t), \widetilde{\beta}_{ij}(t):\nonumber\\ &&V_2<\frac{1}{\delta}\sum_{i=1}^N\sum_{j\in \mathcal{N}_i}\Big(\vartheta\frac{\overline{\alpha}^2}{2}+\chi\frac{\overline{\beta}^2}{2}\Big)\Bigg\},
\end{eqnarray}
where $\delta\leq\min\{\gamma, \mu\vartheta, \nu\chi\}$, {$\overline{\alpha}$ and $\overline{\beta}$ are
two constants,}
\begin{eqnarray}\label{AV 2}
V_2 &=&\xi^T(M\otimes P)\xi\nonumber\\
&&+\sum_{i=1}^N\sum_{j\in \mathcal{N}_i}\bigg(\frac{\widetilde{\alpha}_{ij}(t)^2}{2\mu}{+}\frac{\widetilde{\beta}_{ij}(t)^2}{2\nu}\bigg),
\end{eqnarray}
$\widetilde{\alpha}_{ij}(t)= \alpha_{ij}(t){-}\overline{\alpha} $, $\widetilde{\beta}_{ij}(t)= \beta_{ij}(t){-}\overline{\beta}$, $\overline{\alpha}\geq \frac{1}{2\lambda_{2}}$ and $\overline{\beta}\geq f_0 (N-1)$.
  \item If select $\vartheta$ and $\chi$ small enough, such that $\varrho \triangleq\max\{\mu\vartheta, \nu\chi\}<\gamma$, tracking errors $\xi$ will exponentially converge to the bounded set $\Omega_2$ given as follows:
      \begin{eqnarray}\label{omiga}
\Omega_2\triangleq \bigg\{&&\xi: \|\xi\|\leq\nonumber\\
&&\bigg(\sum_{i=1}^N|\mathcal{N}_i|\frac{\vartheta \overline{\alpha}^2 +\chi \overline{\beta}^2}{2\lambda_{\min}(P)(\gamma-\varrho)}\bigg)^{\frac{1}{2}}\bigg\},
\end{eqnarray}
where $\gamma$ is defined in (\ref{ddddddLV1}).
\end{enumerate}
\end{theorem}
\textbf{Proof}:
Consider the Lyapunov function candidate $V_2$ in (\ref{AV 2}).
As shown in the proof of Theorem \ref{L theorem1f}, the time derivation of $V_2$ along (\ref{A distributed adaptive law}) and (\ref{A closedloop}) satisfies

\begin{eqnarray}\label{dddLV2}
\dot{{ {V}}}_2
&\leq&\xi^T[M\otimes (PA+A^TP)]\xi\nonumber\\
&&-\sum_{i=1}^N\sum_{j\in \mathcal{N}_i}\alpha_{ij}(t)(\xi_i-\xi_j)^T PBB^TP(\xi_i-\xi_j)\nonumber\\
&&+ \sum_{i=1}^N\sum_{j\in \mathcal{N}_i}\bigg( f_0(N-1)\sqrt{N}\|B^TP(\xi_i-\xi_j)\|\nonumber\\
&&- \beta_{ij}(t)\frac{\|B^TP(x_i(t){-}x_j(t))\|^2}{\|B^TP(x_i(t){-}x_j(t))\|{+}\varepsilon e^{-\varphi {t_i}}}\bigg)\nonumber\\
&&{+}\frac{1}{\mu}\sum_{i=1}^N\sum_{j\in \mathcal{N}_i} \widetilde{\alpha}_{ij}(t)\dot{\alpha}_{ij}(t) \nonumber\\
&&{+} \frac{1}{\nu}\sum_{i=1}^N\sum_{j\in \mathcal{N}_i}\widetilde{\beta}_{ij}(t)\dot{\beta}_{ij}(t).
\end{eqnarray}
By using $\Gamma=PBB^TP$, it follows from (\ref{A distributed adaptive law}) that
\begin{eqnarray}\label{A2}
&&-\sum_{i=1}^N\sum_{j\in \mathcal{N}_i}\alpha_{ij}(t)(\xi_i-\xi_j)^T PBB^TP(\xi_i-\xi_j)\nonumber\\
&&+\frac{1}{\mu}\sum_{i=1}^N\sum_{j\in \mathcal{N}_i}  \widetilde{\alpha}_{ij}(t) \dot{\alpha}_{ij}(t)\nonumber\\
&=&{-}2\overline{\alpha} \xi^T(L{\otimes} PBB^TP)\xi{-}\vartheta\sum_{i=1}^N\sum_{j\in \mathcal{N}_i} ( \widetilde{\alpha}_{ij}(t)^2{+}\widetilde{\alpha}_{ij}(t)\overline{\alpha})\nonumber\\
&\leq& {-}2\overline{\alpha} \xi^T(L{\otimes} PBB^TP)\xi\nonumber\\
&&{+}\vartheta\sum_{i=1}^N\sum_{j\in \mathcal{N}_i} \Big(\frac{\overline{\alpha}^2}{2}{-}\frac{\widetilde{\alpha}_{ij}(t)^2}{2}\Big),
\end{eqnarray}
and
\begin{eqnarray}\label{A3}
&&- \sum_{i=1}^N\sum_{j\in \mathcal{N}_i}
\beta_{ij}(t)\frac{\|B^TP(x_i(t)-x_j(t))\|^2}{\|B^TP(x_i(t)-x_j(t))\|+\varepsilon e^{-\varphi t_i}}\nonumber\\
&&+\frac{1}{\nu}\sum_{i=1}^N\sum_{j\in \mathcal{N}_i} \widetilde{\beta}_{ij}(t) \dot{\beta}_{ij}(t)\nonumber\\
&=&-\overline{\beta}\sum_{i=1}^N\sum_{j\in
\mathcal{N}_i}\frac{\|B^TP(x_i(t)-x_j(t))\|^2}{\|B^TP(x_i(t)-x_j(t))\|+\varepsilon e^{-\varphi t_i}}\nonumber\\
&&-\chi\sum_{i=1}^N\sum_{j\in \mathcal{N}_i}(\widetilde{\beta}_{ij}(t)^2+\widetilde{\beta}_{ij}(t)\overline{\beta})\nonumber\\
&\leq& -\overline{\beta}\sum_{i=1}^N\sum_{j\in
\mathcal{N}_i}\frac{\|B^TP(x_i(t)-x_j(t))\|^2}{\|B^TP(x_i(t)-x_j(t))\|+\varepsilon e^{-\varphi t_i}}\nonumber\\
&&+\chi\sum_{i=1}^N\sum_{j\in \mathcal{N}_i}\Big(-\frac{\widetilde{\beta}_{ij}(t)^2}{2}+\frac{\overline{\beta}^2}{2}\Big).
\end{eqnarray}
Substituting (\ref{A2}) and (\ref{A3}) into (\ref{dddLV2}), we have
\begin{eqnarray}\label{dddAV2}
\dot{{ {V}}}_2
&\leq&\xi^T(M {\otimes} (PA+A^TP){-}2\overline{\alpha}L{\otimes} PBB^TP)\xi\nonumber\\
&&+f_0(N-1)\sqrt{N} \sum_{i=1}^N\sum_{j\in \mathcal{N}_i} \|B^TP(\xi_i-\xi_j)\|\nonumber\\
&&-\overline{\beta} \sum_{i=1}^N\sum_{j\in \mathcal{N}_i}
\frac{\|B^TP(x_i(t)-x_j(t))\|^2}{\|B^TP(x_i(t)-x_j(t))\|+\varepsilon e^{-\varphi t_i}}\nonumber\\
&&+\sum_{i=1}^N\sum_{j\in \mathcal{N}_i}\Bigg[ \vartheta\Big(-\frac{ \widetilde{\alpha}_{ij}(t) ^2}{2}+\frac{\overline{\alpha}^2}{2}\Big)\nonumber\\
&&+\chi\Big(-\frac{\widetilde{\beta}_{ij}(t)^2}{2}+\frac{\overline{\beta}^2}{2}\Big)\Bigg].
\end{eqnarray}
As shown in the proof of Theorem \ref{L theorem1f}, by choosing $\overline{\alpha}$ and $\overline{\beta}$ sufficiently large such that  $\overline{\alpha}\geq \frac{1}{2\lambda_{2}}$ and $\overline{\beta}\geq f_0 (N-1)\sqrt{N}$, we have
\begin{eqnarray}\label{AAA v}
\dot{{{V}}}_2 &\leq& -\xi^T(M {\otimes} (PA+A^TP{-} PBB^TP))\xi\nonumber\\
&&+\overline{\beta} \sum_{i=1}^N| \mathcal{N}_i|\varepsilon e^{-\varphi t_i}+\sum_{i=1}^N| \mathcal{N}_i|\Big(\vartheta\frac{\overline{\alpha}^2}{2}+\chi\frac{\overline{\beta}^2}{2}\Big)\nonumber\\
&&-\sum_{i=1}^N\sum_{j\in \mathcal{N}_i}\Bigg( \vartheta\frac{ \widetilde{\alpha}_{ij}(t)^2}{2}+\chi\frac{\widetilde{\beta}_{ij}(t)^2}{2}\Bigg).
\end{eqnarray}
Since $\delta\leq\min\{\gamma, \mu\vartheta, \nu\chi\}$, we obtain that
\begin{eqnarray}\label{AAA vaaa}
\dot{{{V}}}_2&\leq& {-}\delta V_2{+}\sum_{i=1}^N\sum_{j\in \mathcal{N}_i}\frac{(\delta{-}\mu\vartheta) \widetilde{\alpha}_{ij}(t)^2}{2\mu}{+}\frac{(\delta{-}\nu\chi)\widetilde{\beta}_{ij}(t)^2}{2\nu}
\nonumber\\
&&+\overline{\beta}
 \sum_{i=1}^N| \mathcal{N}_i|\varepsilon e^{-\varphi t_i}+\sum_{i=1}^N| \mathcal{N}_i|\Big(\vartheta\frac{\overline{\alpha}^2}{2}+\chi\frac{\overline{\beta}^2}{2}\Big)\nonumber\\
&\leq&-\delta V_2+\overline{\beta}
 \sum_{i=1}^N| \mathcal{N}_i|\varepsilon e^{-\varphi t_i}\nonumber\\
&&+\sum_{i=1}^N| \mathcal{N}_i|\Big(\vartheta\frac{\overline{\alpha}^2}{2}+\chi\frac{\overline{\beta}^2}{2}\Big).
\end{eqnarray}
In light of the well-known Comparison lemma in \cite{kalia}, we can obtain from (\ref{AAA vaaa}) that
\begin{eqnarray}\label{ddddddoV2}
{ {V}}_2(t) &{\leq}& e^{-\delta (t-t_0)}\Big[V_2(t_0)+\frac{1}{\delta}\sum_{i=1}^N\sum_{j\in \mathcal{N}_i}\Big(\vartheta\frac{\overline{\alpha}^2}{2}+\chi\frac{\overline{\beta}^2}{2}\Big)\Big]\nonumber\\
&&{+}\overline{\beta}\sum_{i=1}^N\sum_{j\in \mathcal{N}_i}\int_{t_0}^t\varepsilon e^{-\delta(t-\tau){-}\varphi (\tau+\eta)}d\tau \nonumber\\
&&+\frac{1}{\delta}\sum_{i=1}^N|\mathcal{N}_i|\Big(\vartheta\frac{\overline{\alpha}^2}{2}+\chi\frac{\overline{\beta}^2}{2}\Big).\end{eqnarray}
Therefore, $V_2(t)$ exponentially converges to the bounded set $\Omega_1$ as given in (\ref{omiga1}).
It implies that $\xi(t)$, $\alpha_{ij}(t)$ and $\beta_{ij}(t)$ are uniformly ultimately bounded.

Next, if $\varrho \triangleq\max\{\mu\vartheta, \nu\chi\}<\gamma$, we can obtain a smaller set for $\xi$
by rewriting (\ref{AAA v}) into
\begin{eqnarray}\label{AAA v2}
\dot{{{V}}}_2 &\leq& -\varrho V_2-\lambda_{\min}(P)(\gamma-\varrho)\|\xi\|^2
\nonumber\\
&&+
 \sum_{i=1}^N|\mathcal{N}_i|\bigg[\overline{\beta}\varepsilon e^{-\varphi t_i}{+}\Big(\vartheta\frac{\overline{\alpha}^2}{2}{+}\chi\frac{\overline{\beta}^2}{2}\Big)\bigg].
\end{eqnarray}
Obviously, it follows from (\ref{AAA v2}) that $\dot{V}_2(t)\leq -\varrho V_2(t)+\sum_{i=1}^N|\mathcal{N}_i|\overline{\beta}\varepsilon e^{-\varphi t_i}$, if
$
\|\xi\|^2{>}\frac{\sum_{i{=}1}^N
|\mathcal{N}_i|}{2\lambda_{\min}(P)(\gamma{-}\varrho)}
\Big(\vartheta \overline{\alpha}^2 +\chi \overline{\beta}^2 \Big).
$
Then, in light of $V_2(t)\geq\lambda_{\min}(P)\|\xi\|^2$, we can get that if $\varrho<\gamma$ then $\xi$ exponentially converges to the bounded set $\Omega_2$ in (\ref{omiga}).
Therefore, we obtain from Lemma \ref{A lemma3} that distributed average tracking errors $\xi_i=x_i-\frac{1}{N}\sum_{k=1}^Nr_{k},\;i=1,2,\cdots,N$, converge to the bounded set $\Omega_2$ as $t\rightarrow \infty$. This completes the proof.

\begin{remark}
The adaptive scheme of the algorithm (\ref{A distributed adaptive law}) for updating coupling gains is partly borrowed from  adaptive strategies in \cite{Litac}, \cite{adaptive2}, \cite{adaptive3}, and \cite{adaptive4}. {In Algorithm 1}, it requires the smallest nonzero eigenvalue $\lambda_2$ of $L$, the upper bound $f_0$ of $f_i(t)$ and the number $N$ of nodes in the network. Note that $\lambda_2$, $f_0$ and $N$ are global information for each agent in the network and might not be obtained in real applications. By using adaptive strategies (\ref{A distributed control algorithm}) with (\ref{A distributed adaptive law}) in Theorem \ref{A theorem1f}, the limitation of all these global information can be removed.
\end{remark}
\begin{remark}
Note that related works in \cite{Chengfei:12},  \cite{Chengfei:13}, and \cite{Zhaoyuicca} studied the distributed average tracking problem for integrator-type and linear signals by using non-smooth algorithms, which inevitably produces the chattering phenomenon. Compared with above results, the contribution of this paper is three-fold. First, main results of this paper extend the dynamics from integrator-type signals in \cite{Chengfei:12},  \cite{Chengfei:13} to linear signals. The proposed algorithms (\ref{L distributed control algorithm}) and (\ref{A distributed control algorithm}) successfully solve the distributed average tracking problem for reference signals generated by the more general linear dynamics. Second, by using adaptive control approaches, the limitation of all these global information is removed. Third, compared with existing results in \cite{Zhaoyuicca}, new continuous algorithms are redesigned via the boundary layer concept, which plays a vital role to reduce the chattering phenomenon in real applications.
\end{remark}

\section{Simulations}
In this section, we will give an example to verify Theorem \ref{A theorem1f}.  The dynamics of multiple reference signals are given by (\ref{L referencesignals}) with
$
r_i=\left(
  \begin{array}{c}
     {r}_{1i} \\
     {r}_{2i} \\
  \end{array}
\right),\;\;A=\left(
          \begin{array}{cc}
            0 & 1 \\
            -1 & -2 \\
          \end{array}
        \right),\;\;B=\left(
  \begin{array}{c}
    0 \\
    1 \\
  \end{array}
\right), \nonumber
$
and $f_i(t)=\frac{i+1}{2} \sin(t)$, where $i=1,2,\cdots, 6$.
The communication topology is shown in Fig. 1.
Solving the ARE (\ref{LMI}) with $Q=I$ gives
the gain matrices $K$ and $\Gamma$ as
$
K{=}\left(
  \begin{array}{cc}
    -1.5728 \; &  -4.3293 \\
  \end{array}
\right), \; \Gamma{=}
\left(
  \begin{array}{cc}
    2.4738 \; &  6.8092\\
    6.8092 \; & 18.7428\\
  \end{array}
\right).\nonumber
$
The state trajectories $x_i, \; i=1,2,\cdots,6$, of six agents under Algorithm 2 with $\mu=10,\;\nu=10,\;\vartheta=0.01,\;\chi=0.01,\;\varepsilon=5,\;\varphi=0.5,\; K$ and $\Gamma$ given above are depicted in Fig. 2, which shows that {states achieve a small bounded neighborhood of the average value of all signals.} It follows from Fig. 3 that tracking errors $\xi_i\triangleq  x_i-\frac{1}{6}\sum_{k=1}^6r_k$ convergent  to  {a small bounded neighborhood of the origin} as $t\rightarrow \infty$. The adaptive coupling gains $\alpha_{ij}(t)$ and $\beta_{ij}(t)$ are also drawn in Fig. 4, respectively. { As a comparison, the discontinuous algorithm in \cite{Zhaoyuicca} and continuous algorithms (\ref{A distributed control algorithm}) are also shown with the same parameters in Fig. 5, where we can see that the chattering  effect with discontinuous algorithm in \cite{Zhaoyuicca} is greatly reduced by using the continuous algorithm (\ref{A distributed control algorithm}).}

\begin{figure}
  \center{
  \includegraphics[width=3cm]{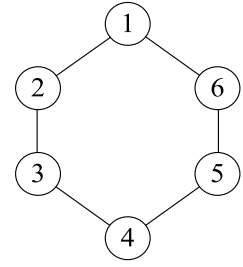}\\
  \caption{The communication topology.} }
\end{figure}

\begin{figure}
  \center{
  \includegraphics[width=6cm]{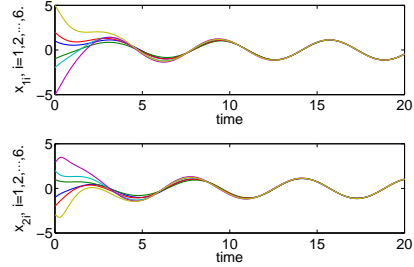}\\
  \caption{State trajectories $x_i$ of six agents in networks.} }
\end{figure}

\begin{figure}
  \center{
  \includegraphics[width=6cm]{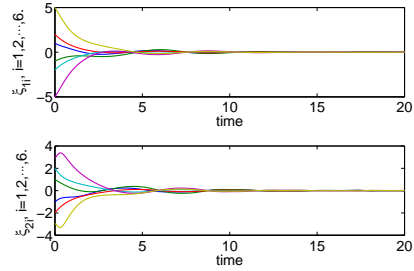}\\
  \caption{Tracking error trajectories $\xi_i=x_i-\frac{1}{6}\sum_{k=1}^6r_k$ of  six agents in the network.} }
\end{figure}

\begin{figure}
  \center{
  \includegraphics[width=6cm]{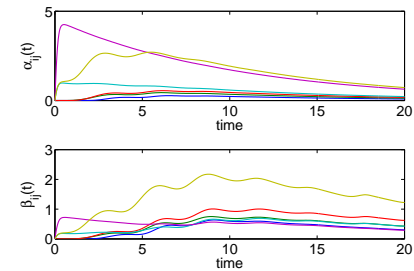}\\
  \caption{Adaptive coupling strengths $\alpha_{ij}(t)$ and $\beta_{ij}(t)$ in (\ref{A distributed adaptive law}).} }
\end{figure}

\begin{figure}
  \center{
  \includegraphics[width=6cm]{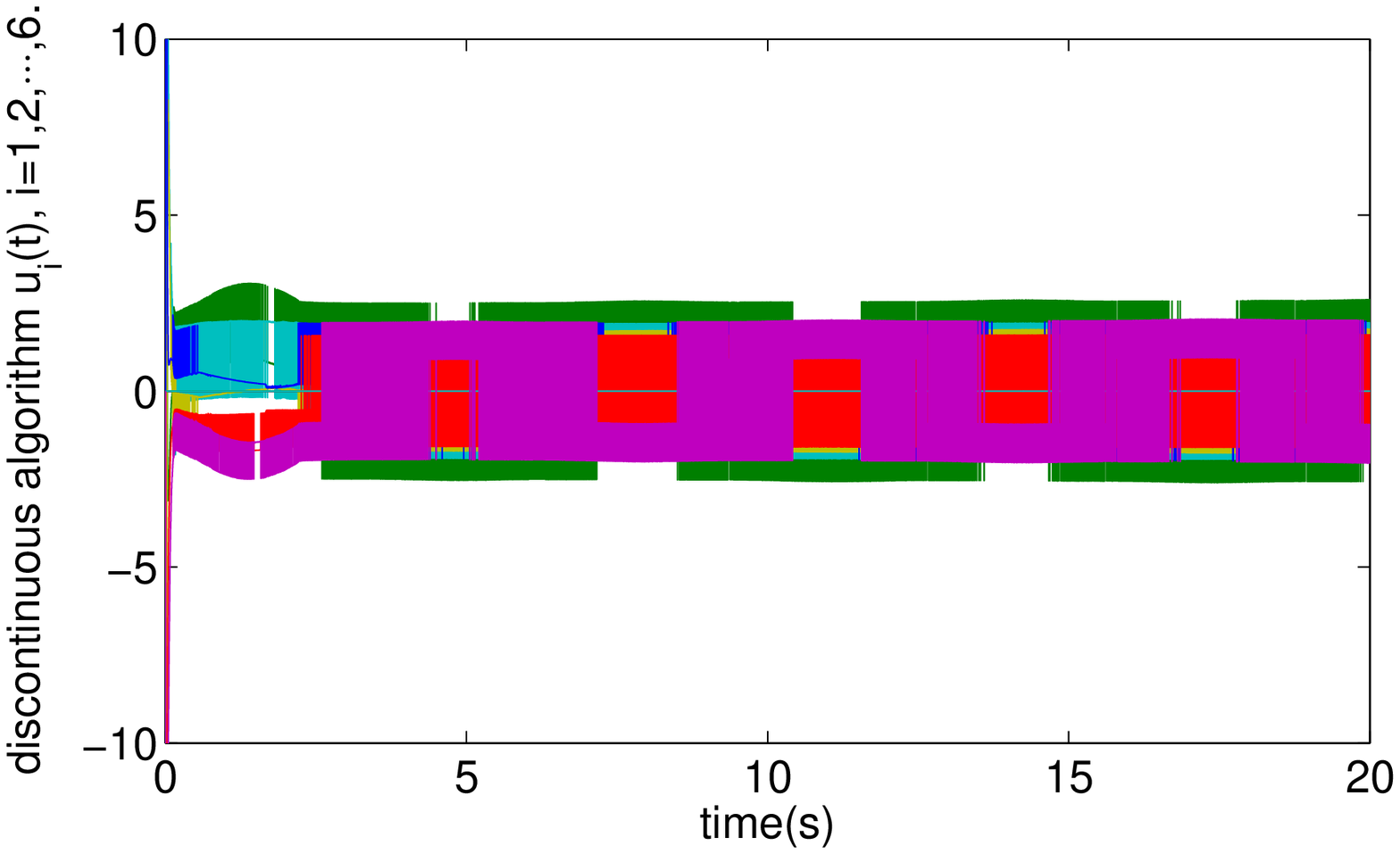}\\
  \includegraphics[width=6cm]{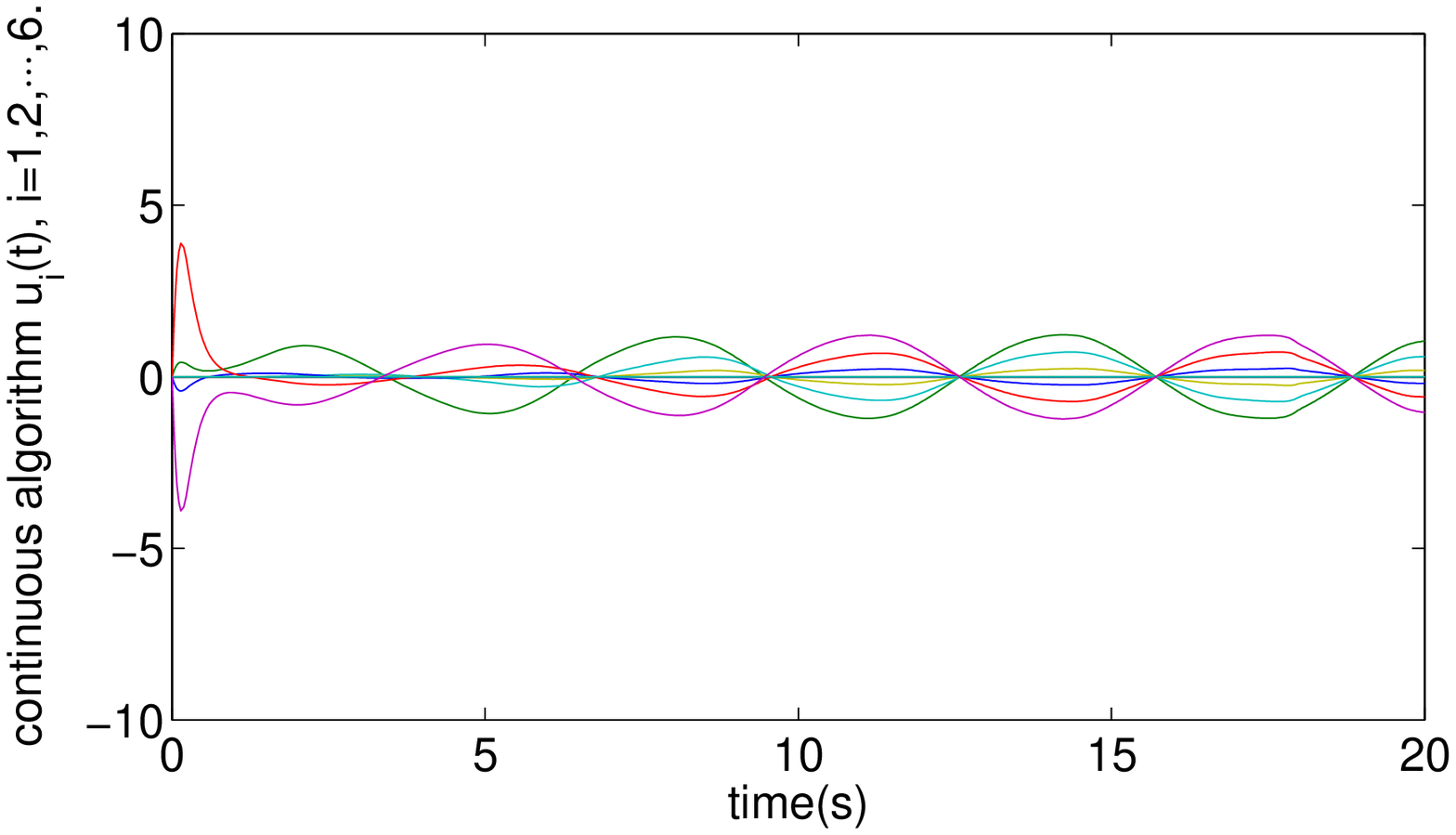}\\
  \caption{Discontinuous algorithm in \cite{Zhaoyuicca} and the continuous algorithm (\ref{A distributed control algorithm}).} }
\end{figure}
\section{Conclusions}
In this paper, we have studied the distributed average tracking problem of multiple time-varying signals generated by general linear dynamical systems, whose reference inputs are nonzero, bounded and not available to any agents in networks. In {the} distributed fashion, a pair of continuous algorithms with static and adaptive coupling strengths have been developed in light of the boundary layer concept. Besides, sufficient conditions for the existence of distributed algorithms are given if each agent is stabilizable. The future topic will be focused on the distributed average tracking problem for the case with only the relative output information of neighboring agents.

\bibliographystyle{plain}        


\end{document}